\documentclass[graybox]{svmult}
\usepackage[]{graphicx}
\usepackage{graphics}
\usepackage{amssymb}
\usepackage{amsmath}
\usepackage[hyperindex=true,pdftitle={Characterizing price index behavior
through fluctuation dynamics},pdfauthor={Sayantan Ghosh},colorlinks=true,
linkcolor=black,breaklinks=true,citecolor=blue,plainpages=false,
pdfpagelabels]{hyperref}
\usepackage{pstricks}
\usepackage{subfigure}
\usepackage{multirow}
\usepackage{color}
\usepackage{mathptmx}       
\usepackage{helvet}         
\usepackage{courier}        
\usepackage{type1cm}        
\usepackage[round,numbers,sort&compress]{natbib}
\usepackage{makeidx}         
\usepackage{graphicx}        
\usepackage{multicol}        
\usepackage[bottom]{footmisc}
\usepackage{url}
\usepackage[top=0.75in, bottom=0.75in, left=1in, right=1in]{geometry}

\makeindex             

\begin{document}

\title*{Characterizing price index behavior through fluctuation dynamics}
\author{Prasanta K. Panigrahi$^{1*}$, Sayantan Ghosh$^{1,2\ddagger}$, Arjun
Banerjee$^{3}$, Jainendra Bahadur$^{4}$ and P. Manimaran$^{5\dagger}$}
\authorrunning{Panigrahi et al.}
\institute{$^1$Dept. of Physical Sciences, Indian Institute of Science Education
and Research Kolkata, BCKV Campus, PO Mohanpur, Dist. Nadia, West Bengal 741
252, India
\and $^*$\email{pprasanta@iiserkol.ac.in} 
\and $^2$Center for Quantum Technology, School of Physics, University of
KwaZulu-Natal, Private Bag X54001, Durban 4000, South Africa 
\and $^{\ddagger}$\email{210556397@ukzn.ac.za, say.xen@gmail.com}
\and $^3$Dept. of Metallurgical and Materials Engineering, National Institute of
Technology, Durgapur 713 209, West Bengal, India
\and $^4$ Dept. of Electronics and Communication Engineering, National Institute
of Technology, Durgapur 713 209, West Bengal, India
\and $^4$C R Rao Advanced Institute of Mathematics, Statistics and Computer
Science, Gachi Bowli, Hyderabad 500 046, India 
\and $^{\dagger}$\email{rpmanimaran@gmail.com}
}
%
%
\maketitle

\abstract*{We study the nature of fluctuations in variety of price indices
involving companies listed on the New York Stock Exchange. The fluctuations at
multiple scales are extracted through the use of wavelets belonging to
Daubechies basis. The fact that these basis sets satisfy vanishing moments
conditions makes them ideal to extract local polynomial trends, through the low
pass or `average coefficients'. Subtracting the trends from the original time
series yields the fluctuations, at different scales, depending on the level of
low-pass coefficients used for finding the `average behavior'. The fluctuations
are then studied using wavelet based multifractal detrended fluctuation analysis
to analyze their self-similar and non-statistical properties. Due to the
multifractality of such time series, they deviate from Gaussian behavior in
different frequency regimes. Their departure from random matrix theory
predictions in such regimes is also analyzed. These deviations and
non-statistical properties of the fluctuations can be instrumental in throwing
significant light on the dynamics of financial markets.
}

\abstract{We study the nature of fluctuations in variety of price indices
involving companies listed on the New York Stock Exchange. The fluctuations at
multiple scales are extracted through the use of wavelets belonging to
Daubechies basis. The fact that these basis sets satisfy vanishing moments
conditions makes them ideal to extract local polynomial trends, through the low
pass or `average coefficients'. Subtracting the trends from the original time
series yields the fluctuations, at different scales, depending on the level of
low-pass coefficients used for finding the `average behavior'. The fluctuations
are then studied using wavelet based multifractal detrended fluctuation analysis
to analyze their self-similar and non-statistical properties. Due to the
multifractality of such time series, they deviate from Gaussian behavior in
different frequency regimes. Their departure from random matrix theory
predictions in such regimes is also analyzed. These deviations and
non-statistical properties of the fluctuations can be instrumental in throwing
significant light on the dynamics of financial markets.
}

\section{Introduction}
\label{sec:1}
Financial time-series which were in the past of interest to only economists,
have led to considerable inter-disciplinary research due to the applicability of
various physical laws and techniques in their analysis. This has led to the
discovery of various new aspects like fractality \cite{mandelbrot1963},
multifractality \cite{peng1994}, correlated behavior
\cite{bouchad2000,mantegna2000} and complex network structure
\cite{mantegna1999hier}. The last few years has also seen a lot of activity in
terms of explaining the correlations in financial markets through the Random
Matrix Theory (RMT) framework \cite{wigner1951,plerou1999,plerou2002}. 
\par
Fractals as first predicted by Benoit Mandelbrot in the 1960s
\cite{mandelbrot1982} have been widely applied to understand various processes
in Physics \cite{tosatti1986}, Chemistry \cite{stanleynature1988} and Biology
\cite{dewey1997fractals}. Mandelbrot in 1963 proposed the study of fluctuations
in the market prices \cite{mandelbrot1963} which opened new vistas for the
analysis of stock markets through statistical physics. In the recent years, Peng
\textit{et al.} proposed the ``Detrended Fluctuation Analysis (DFA)'' in 1994
\cite{peng1994} to study the DNA nucleotide structure as a random walk problem
which was extended to study the price fluctuations in economic time-series under
a mono-fractal hypothesis \cite{liu1999,vandewalle1999,kantelhardt2001}.
However, the inadequacy of the mono-fractal hypothesis to model the behavior of
financial time series was soon pointed out and consequently, a multi-fractal
model called the ``Multi-Fractal Detrended Fluctuation Analysis (MFDFA)'' was
proposed \cite{kantelhardt2002a}. This method used a variable window approach to
calculate the local variances in the profile of the data series from the
polynomial fit. Manimaran \textit{et al.} in 2009\cite{mani3} building on the
work by Kantelhardt \textit{et al.} proposed the ``Wavelet Based Multi-Fractal
De-trended Fluctuation Analysis (WBMFDFA)'', where, using the Multi-Resolution
Analysis (MRA) capable ``fractal like'' kernels, the time-frequency resolution
and extraction of fluctuations for multi-fractal analysis was shown to have a
greater efficiency than its predecessor. 
\par
The study of the correlation matrix of the financial return series have been
shown to agree well with the predictions of RMT and the nearest-neighbor-spacing
of the rank-ordered unfolded eigenvalues of the correlation matrix follow that
of the Gaussian Orthogonal Ensemble (GOE) \cite{laloux1999a,plerou1999}. This
behavior dubbed as an \textit{universal} behavior of the financial return series
have again, recently been studied with respect to temporal evolution of
financial correlations to study the differences between the assumption of
strongly correlated financial time-series and uncorrelated financial time-series
\cite{fenn2011}. In this context, it becomes important to analyze the nature of
correlations in the time-frequency domain in order to ascertain the effects of
non-stationarity and transience on such studies. The inefficient handling of
such signals by either Fourier Transform or Short Term Fourier Transform for the
purpose of a time-frequency localized study have already been established,
leading to the development of Wavelet Transform \cite{daub1}. 
\par
In this work, we will use a wavelet based fluctuation extraction technique to
study the correlations of the fluctuations at various frequency windows (called
\textit{scales} in the wavelet parlance). We will also briefly comment on the
multi-fractal nature of the fluctuations and the distributions of the associated
parameters: Hurst exponent and singularity strength. The organization of the
article is as follows: in Sec. (\ref{sec:review}), we will briefly review the
theoretical methods of Wavelet Based Fluctuation Extraction (WBFE), Wavelet
Based Multi-Fractal De-trended Fluctuation Analysis (WBMFDFA) and the Random
Matrix Theory (RMT) based method to study the time-frequency localized
correlations of the fluctuations. Further in Sec. (\ref{sec:RND}), we discuss
the results obtained by the application of WBFE, WBMFDFA and correlation
analysis on the price index of $196$ scrips trading on the New York Stock
Exchange (NYSE) between September 1984 to June 2010. Finally, we summarize and
conclude with the scope for future work in Sec. (\ref{sec:conc}).
\section{Review of theoretical methods}
\label{sec:review}
\subsection{Wavelet Based Fluctuation Analysis}
In the following, the analysis of a time series given by $X(t)$ is carried out
by calculating the ``log-normalized return series'' $R(t)$:
\begin{eqnarray}
r(t)&=&\log X(t+1)-\log X(t)\\
R(t)&=&\frac{r(t)-\langle r(t)\rangle}{\sigma_r}
\end{eqnarray}
where, $\langle\cdot\rangle$ and $\sigma_r$ are the time average and standard
deviation of the log-return series $r(t)$. $\sigma_r$ is also called as the
``volatility of returns''. The profile $Y(t)$ is calculated by taking the
cumulative sum of the log-normalized return series:
\begin{equation}
Y(t)=\sum\limits_{k=1}^{t} R(k)
\end{equation}
which is then subjected to the Wavelet Based Fluctuation Extraction
(WBFE)\cite{mani3}. The WBFE can be performed following the steps:
\begin{enumerate}
\item Calculate the one-dimensional discrete wavelet transform
(1DWT)\cite{mallat1,daub1,farge1,torrence1} of the profile:
\begin{equation}
Y(t)=\sum\limits_{b=-\infty}^{\infty} c_b \phi_b(t)+\sum\limits_{a\geq
0}^{l}\sum\limits_{b=-\infty}^{\infty}d_{ab}\psi_{ab}(t)
\end{equation}
where $c_b$ are the ``low-pass'' coefficients that capture the trend or the
average behavior of the signal and $d_{ab}$ are the ``high-pass'' coefficients
capturing the local fluctuations in the signal at various window sizes $a$. The
functions $\Phi$ and $\Psi$ are called the ``scaling filter'' and the ``high
pass'' filters respectively. The father and mother wavelets $\phi_b(t)$ and
$\psi_{ab}(t)$ are orthogonal to each other and are subjected to the
admissibility conditions
\begin{eqnarray}
\int \phi(t)dt &<&\infty\\
\int \psi(t)dt &=&0\\
\int \phi^*(t)\psi(t)dt&=&0,\mbox{ orthogonality}\\
\int \vert\phi(t)\vert^2dt=&1&= \int \vert\psi(t)\vert^2dt
\end{eqnarray}
The scaled and translated versions of $\psi(t)$ are called the ``daughter
wavelets''
\begin{equation}
\psi_{ab}(t)=2^{a/2}\psi(2^at-b), a\in \mathbb{R}, b\in \mathbb{Z}^+
\end{equation}
which differ from the mother wavelet $\psi(t)$ at the $a^{th}$ scale by $2^a$ in
height and $2^{a/2}$ in width. $a$ and $b$ are called the scaling and
translation parameters respectively and $l=\lfloor \log N/\log 2 \rfloor$ is the
maximum number of scales for the profile $Y(t)$ of length $N$.
\par 
The wavelet kernel for the 1DWT should be chosen such that it captures the
maximum information from the signal. For example, the Daubechies' family of
wavelets satisfy vanishing moment conditions which make them blind to various
polynomial trends. The wavelet Db-$N$ (with the index number $N$ being even
integers between 2 and 20), has $N/2$ vanishing moments limiting the
representation of a polynomial trend of $N/2$ in the signal. The Db-$4$ wavelet
has two vanishing moments making it blind to constant and linear trends. In this
work the Db-$4$ wavelet is employed. 
\item Calculate the approximate trend $T_a(t)$ at the scale of interest $a$ and
subtract it from the profile $Y(t)$ to get the fluctuations $Z_a(t)$: 
\begin{equation}
Z_a(t)=Y(t)-T_a(t).
\label{eq:flucWB}
\end{equation}
\end{enumerate}
The $Z_a(t)$ obtained by this method represent the fluctuations at different
frequency bands (the scale is inversely related to the frequency). Consequently,
these fluctuations can be probed to analyze the behavior of the signal in
various frequency bands like Fourier power law and moments of the fluctuation
distribution. It has been shown earlier that the well-known $f^{-3}$ behavior of
market fluctuations appear only in the low frequency or long wavelength
regime\cite{sayantan2011}. Due to the varying window sizes (corresponding to
different scales) and the convolution error generic to wavelet transforms, the
extracted fluctuations happen to have erroneous values at the edges. These
errors are corrected by performing the WBFE on the reversed profile and then
taking the average of the two (forward and reversed) fluctuation series at each
scale. 
\subsection{Wavelet Based Multi-Fractal De-trended Fluctuation Analysis}
These fluctuations can also be subjected to a multi-fractal analysis which is a
modified form of the original MFDFA proposed by Kantelhardt \textit{et al.} in
2002\cite{kantelhardt2002a}. 
\par
The fluctuations obtained in Eq. \eqref{eq:flucWB} are further subdivided into
$N_s=\lfloor N/s \rfloor$ segments of size $s$ such that $s=2^{a-1}W$, where $W$
is the support width of the wavelet and $a$ represents the scale. Thus the
fluctuations obtained at various scales can be analyzed at window sizes
corresponding to the scale and the wavelet used. 
\par
Since the fluctuations are guaranteed to have zero mean, we can directly find
the variance of each segment and thus calculate the fluctuation function
\begin{equation}
F_q(s)=\left[\frac{1}{2N_s}\sum\limits_{k=1}^{s}\left\{\sigma^2(k,s)\right\}^{
q/2}\right]^{1/q},\qquad q\neq 0
\end{equation}
$F_q(s)$ is the ``$q$th order'' fluctuation function, where $q\in[-m,m],~ m\in
\mathbb{Z}^+$. The negative (positive) $q$ values capture the fractality of the
broader (finer) fluctuations. It can be easily seen that at $q=0$, the $F_0(s)$
blows up. Hence, to circumvent this issue, at $q=0$, 
\begin{equation}
F_0(s)=\exp\left[\frac{1}{2N_s}\sum\limits_{k=1}^{s}\log\left\{\sigma^2(k,
s)\right\}^{q/2}\right]^{1/q},\qquad q=0.
\end{equation}
The ``generalized Hurst exponent'' $h(q)$ can be obtained from $F_q(s)$ since,
as a function of the segment size $s$, $F_q(s)$ follows a power law of the form
\begin{equation}
F_q(s)\sim s^{h(q)}.
\end{equation}
It must be noted that at $q=2$, this method reduces to the standard mono-fractal
fluctuation analysis and $h(2)$ is the Hurst exponent. The multi-fractal scaling
exponent $\tau(q)$ can be calculated as
\begin{equation}
\tau(q)=qh(q)-1
\end{equation}
The singularity spectrum $f(\beta)$ is related to the multi-fractal scaling
exponent $\tau(q)$ by a Legendre transform
\begin{equation}
\beta=\frac{d}{dq}\tau(q),\quad\mbox{and}\quad
f\left(\frac{d}{dq}\tau(q)\right)=q\frac{d}{dq}\tau(q)-\tau(q)\equiv
f(\beta)=q\beta-\tau(q)
\end{equation}
\subsection{Correlation analysis of fluctuations}
The fluctuations $Z_a$ obtained at scale $a$ through Eq. \eqref{eq:flucWB} can
be analyzed for correlations. The fluctuations for the entire data set of $N$
scrips can be written as a $N\times T$ matrix $\mathcal{X}$, where $T$ is the
time length of the fluctuation series of each the $N$ scrips. The correlation
matrix is thus given by
\begin{equation}
\mathcal{C}(x,y)=\frac{1}{T}\mathcal{X}\mathcal{X}^T,
\end{equation}
where, $\cdots^T$ is the transposition operation. In the case where
$\mathcal{X}$ consisted of $N$ mutually independent normally distributed
fluctuations of length $T$, $\mathcal{C}$ could be considered to be a Wishart
matrix \cite{edelman1988,baker1998,laloux1999a}. Under the constraint
$N\rightarrow\infty,T\rightarrow\infty$ and $Q\equiv T/N \geq 1$ the density of
eigenvalues of the correlation matrix takes the form \cite{mitra1999}
\begin{equation}
\rho(\Lambda)=\frac{Q}{2\pi\sigma_\mathcal{X}^2}\frac{\sqrt{(\Lambda_{\textrm{
max}}-\Lambda)(\Lambda-\Lambda_{\textrm{min}})}}{\Lambda},
\label{eq:eqrholamb}
\end{equation}
with $\sigma_{\mathcal{X}}^2$ being the variance of the matrix $\mathcal{X}$ and
$\Lambda^{\textrm{max}}_{\textrm{min}}$ given by
\begin{equation}
\Lambda^{\textrm{max}}_{\textrm{min}}=\sigma_\mathcal{X}^2\left[1+\frac{1}{Q}
\pm2\sqrt{\frac{1}{Q}}\right].
\end{equation}
It must be noted that under the constraint $N\rightarrow\infty$, the eigenvalues
of the the matrix $\mathcal{C}$ lie strictly in the range
$[\Lambda_{\textrm{min}},\Lambda_{\textrm{min}}]$. However, for finite sized
matrices, there exists a finite probability of finding eigenvalues outside this
range. Indeed, it has been shown that for financial time series, the
nearest-neighbor eigenvalue spacing $\lambda\equiv\Lambda_{i+k}-\Lambda_i$,
obtained through unfolding the eigenvalues of $\mathcal{C}$ follow the
distribution for a Gaussian Orthogonal Ensemble (GOE)\cite{plerou1999}
\begin{equation}
\rho(\lambda)=\frac{\pi \lambda}{4}\exp\left(-\frac{\pi}{2}\lambda^2\right).
\end{equation}
Here, we will investigate the scale dependence of the $\rho(\lambda)$ for the
correlation of the fluctuations $Z_a(t)$ and $Z'_a(t)$ at scale $a$, where
$Z_a(t)$ and $Z'_a(t)$ are different scrips in the corpus. This will give us an
idea about the effectiveness of the RMT predictions in different frequency
regimes.
\section{Results and discussion}
\label{sec:RND}
\subsection{Data}
We have analyzed 196 scrips trading on the New York Stock Exchange (hereafter
referred to as NYSE) in the period from September the 7th, 1984 to June the
10th, 2010. The scrips analyzed in this work have been chosen so as to encompass
high-cap, mid-cap as well as low-cap sections of the American stock market. This
combined with the time-frame for analysis includes some of the major crashes of
the NYSE, for the example the \textit{Black Monday} on October 1s9, 1987, the
July 2, 1997 crash triggered by the Asian financial crisis, the burst of the
\textit{dot-com bubble} leading to a three years sluggish activities from March
10, 2000 and the two years long bear run of the market from 2007-2009. The data
has been analyzed using the WBMFDFA method and also through a correlation
analysis of the fluctuations in the random matrix theory framework. 
\subsection{Time-frequency localized correlation analysis}
\begin{figure}
\centering
\subfigure[]{\label{fig:rholamb}\includegraphics[width=0.4\textwidth]{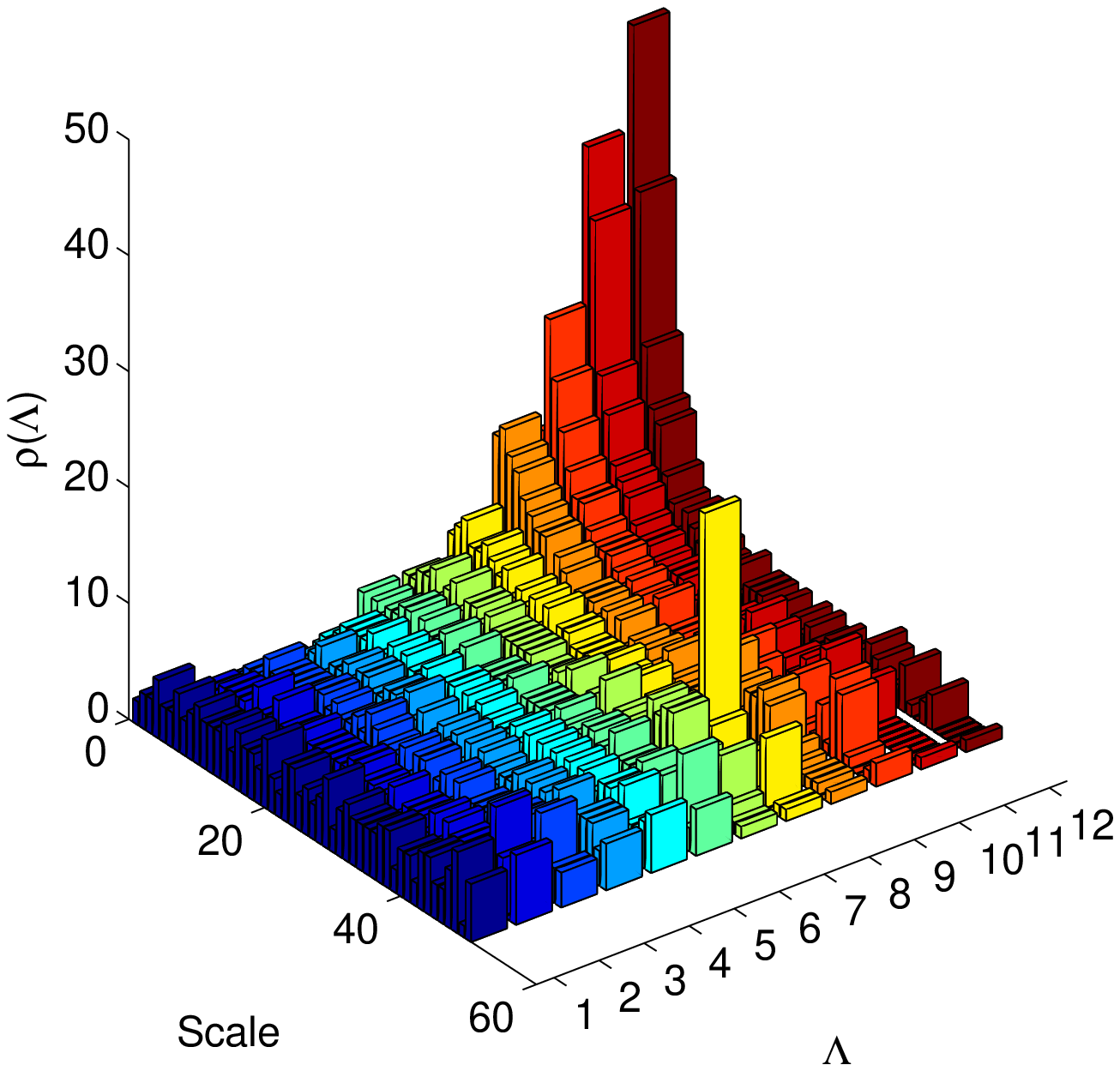}}
\subfigure[]{\label{fig:fitrholamb}\includegraphics[width=0.4\textwidth]{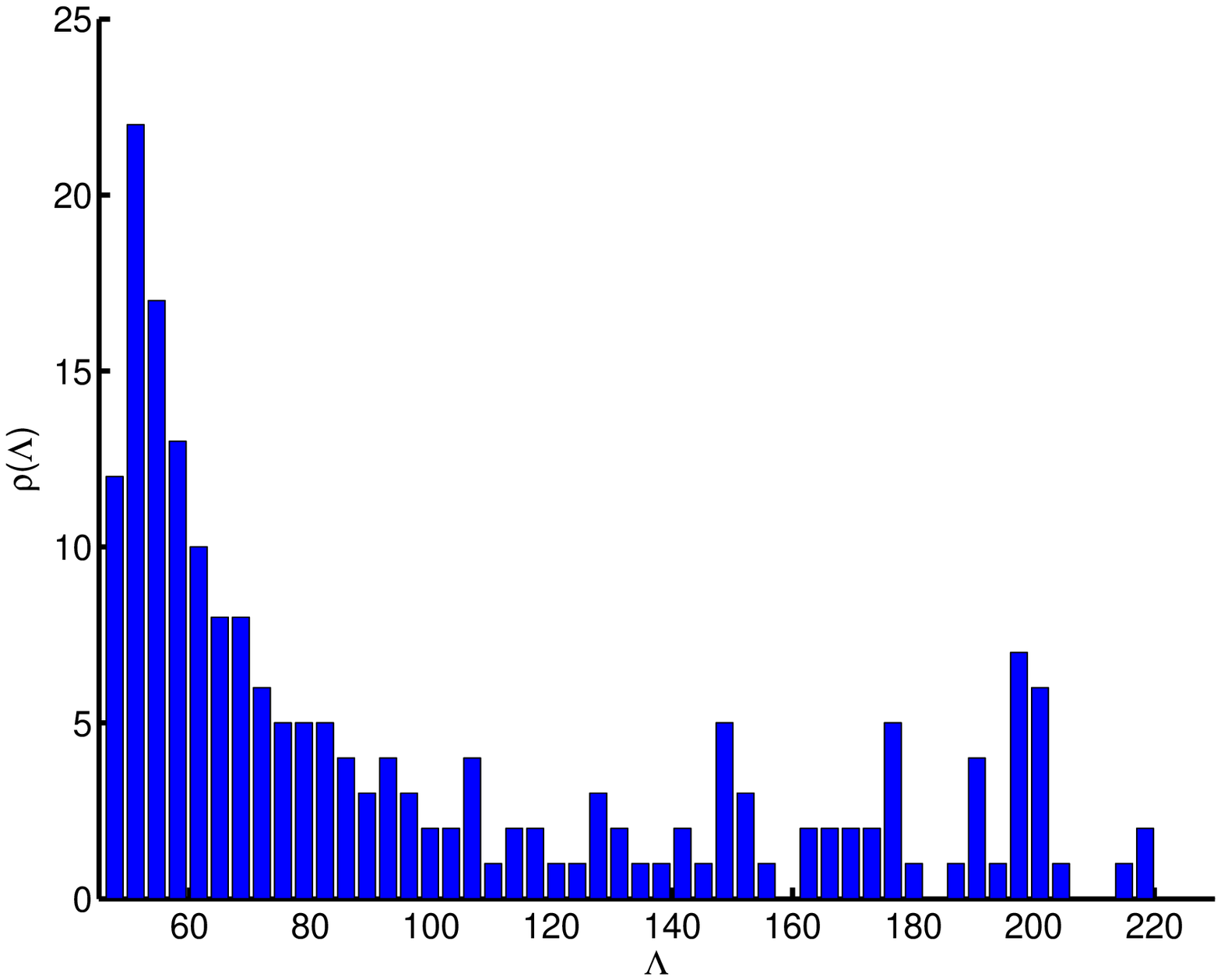
}}
\caption{\label{fig:denserholamb}Plots depicting the density of unfolded
eigenvalues $\rho(\Lambda)$ of the correlation matrix of the fluctuations at
\subref{fig:rholamb} different scales and at \subref{fig:fitrholamb} scale $10$.
It can be seen the $\rho(\Lambda)$ fits with the analytical Eq.
(\ref{eq:eqrholamb}).}
\end{figure}
Analysis and investigation of correlations of the fluctuations over different
frequency windows $a$ ($a\propto 1/f$ where $f$ is the frequency) can provide us
with insight into the spectral behavior of the market correlations thereby
improving our understanding of the collective behavior of the market in
different time spans. For example, if the correlations between different scrips
representing different sectors of the market in short time windows (high
frequency, low scale) is low, then this could be exploited to guard the
simultaneous crashes of different sectors in the event of a crisis. However, if
the scrips are correlated in the low frequency limit, then the long term
correlations of the market could lead to the ``healing'' of the market in a
systematic and efficient way after a crash. This kind of information could be
very useful for the policy makers in order to both monitor the economy as well
provide for safe-guards against possible crashes.
\par
\begin{figure}
\centering
\subfigure[]{\label{fig:histeig}\includegraphics[width=0.4\textwidth]{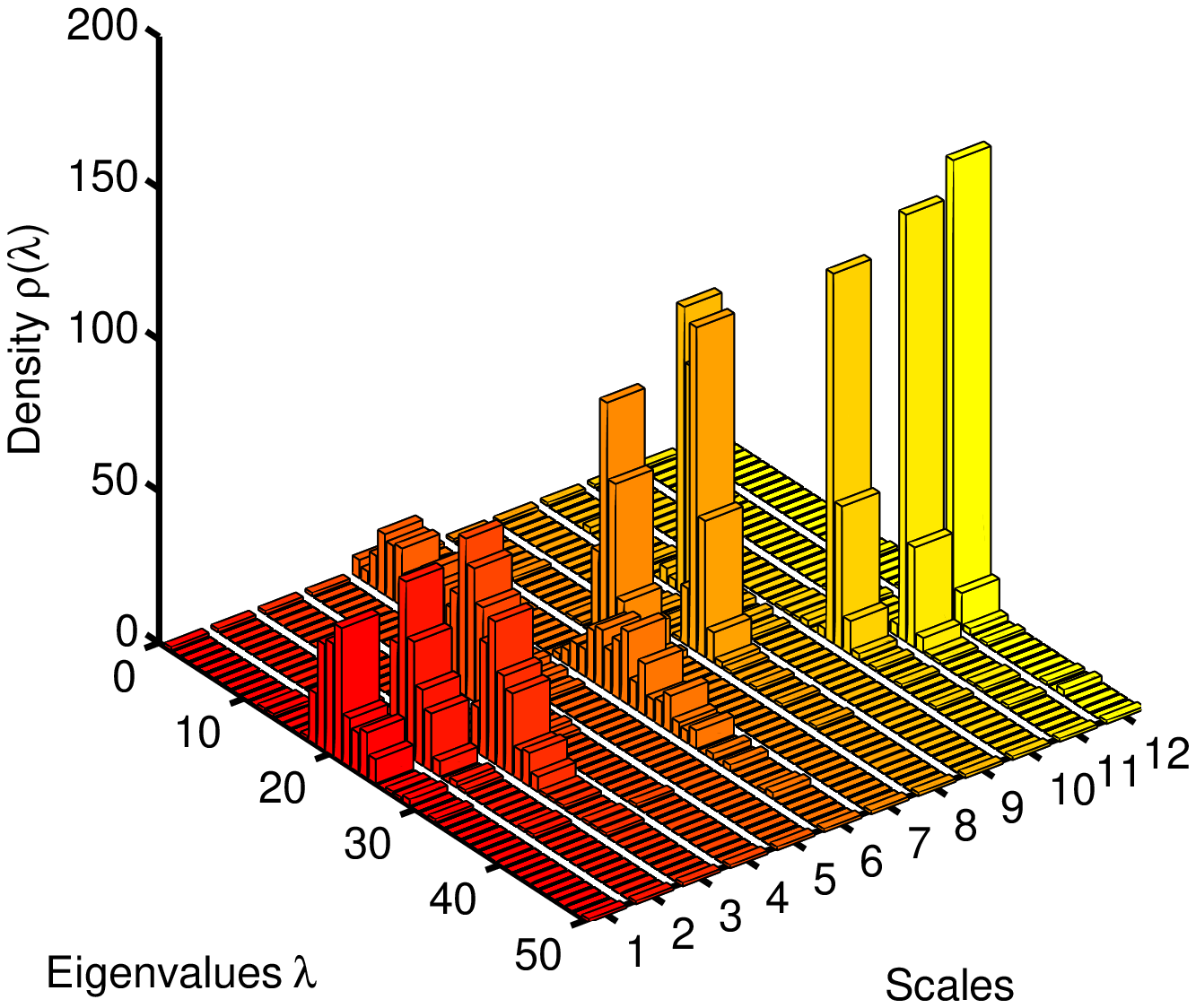}}
\subfigure[]{\label{fig:fithisteig}\includegraphics[width=0.4\textwidth]{
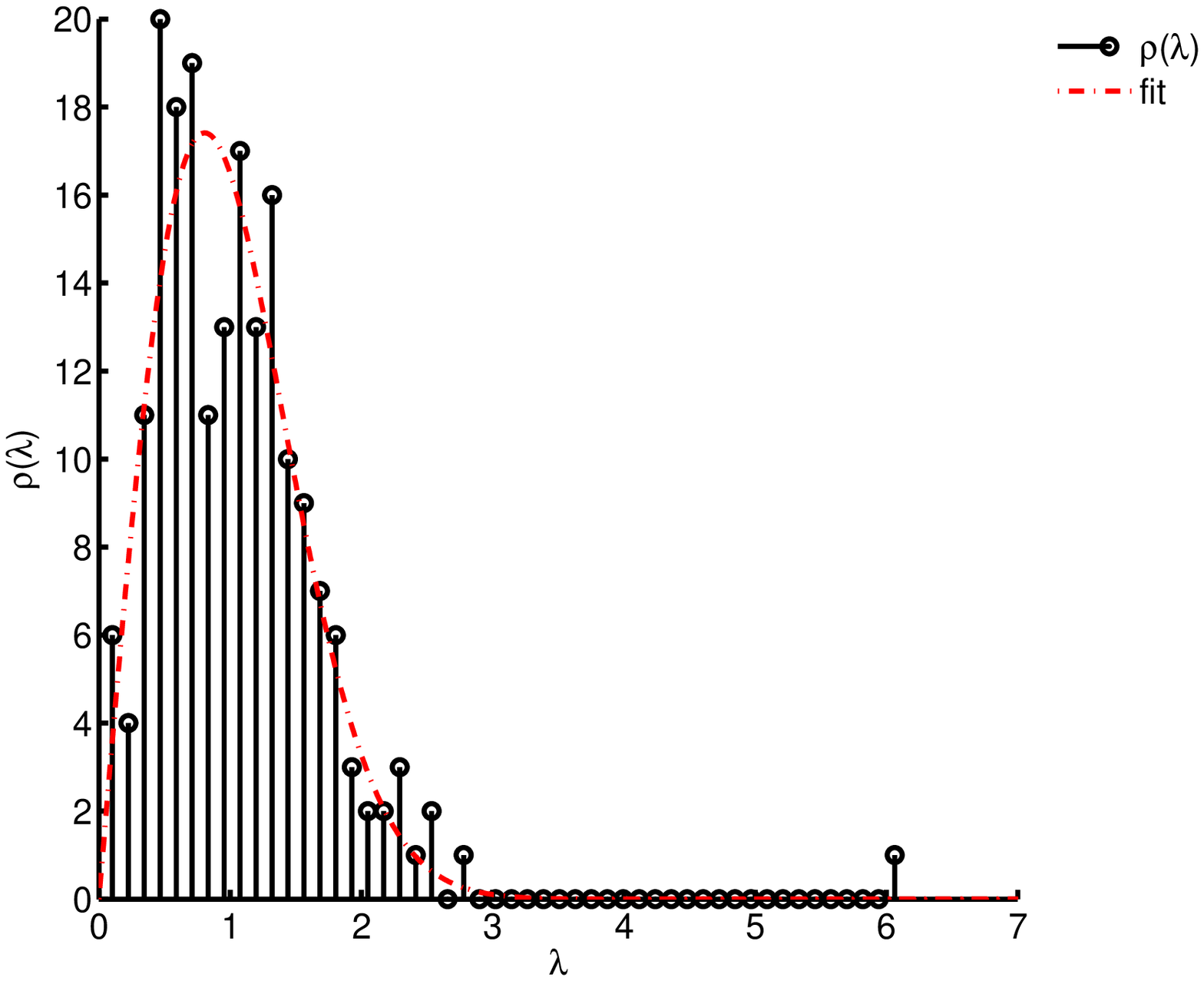}}
\caption{\label{fig:figeigs}Plots depicting the density of nearest neighbor
unfolded eigenvalue spacing $\rho(\lambda)$ of the correlation matrix of the
fluctuations at \subref{fig:histeig} different scales and at
\subref{fig:fithisteig} scale $5$, $\rho_5(\lambda)$ fit with
$\rho(\lambda)=a\lambda\exp\left(-b\lambda^2\right)$ where $a=35.64$ and
$b=0.7707$ (with $95\%$ confidence bounds) as an illustration. We can see that
though the GOE fits well in the high and mid frequency ranges, at low frequency
or long periods, they do not fit with the RMT assumptions.}
\end{figure}
In order to understand the nature of spectral correlations in the market, we
analyze the spectrum of the correlation matrices at different scales. In Fig.
(\ref{fig:denserholamb}), we have shown the density of the unfolded eigenvalues
of the correlation matrices of the fluctuations obtained at different scales. We
observe from Fig. (\ref{fig:rholamb}) that at lower and middle scales (upto
$a=7$), the unfolded eigenvalue distribution $\rho(\Lambda)$ is largely uniform
and changes to fit the Eq. (\ref{eq:eqrholamb}) at higher scales as shown in
Fig. (\ref{fig:fitrholamb}). Since in this analysis, we have looked at the
behavior of the fluctuations over the whole time period $T=5799$, the value of
$Q=29.58$ is very large. We could expect that the correlations of such
fluctuations will show significant deviations from the RMT prediction since it
fits well in short time windows\cite{kumar2012}. We must remember that the
fluctuations obtained by the WBFE method at characteristic of the frequency
range (scale) under study. The spectral correlations under investigation here
show that at high frequencies (low scales), the GOE fit of Eq.
(\ref{eq:eqrholamb}) is not followed by $\rho(\Lambda)$. However, at higher
scales, that is at lower frequencies, the $\rho(\Lambda)$ fits reasonably well
with the Eq. (\ref{eq:eqrholamb}).
\par
In Fig. \ref{fig:histeig}, we have plotted the density of the eigenvalue spacing
$\rho(\lambda)$ against different scales. It is clearly visible that at the
lower and middle scales, the $\rho(\lambda)$ follows the RMT prediction well,
which we have exemplified in Fig. \ref{fig:fithisteig} at scale $5$. This is in
contrast with the results in Fig. (\ref{fig:rholamb}) suggesting that the
eigenvalue spacing at lower scales is behaves differently from the eigenvalues
themselves. This is an interesting observation which we believe should be
explored in greater details.

\section{Conclusion}
To summarize, we have analyzed the nature of fluctuations from different sectors
of the New York Stock Exchange (NYSE) through Wavelet based multi-fractal
analysis and RMT based techniques. The number of companies being small, it is
expected that there will be significant deviation from the RMT prediction.
Interestingly, although the density of unfolded eigenvalues exhibited this
behavior, the nearest-neighbor eigenvalue spacing distribution matched
reasonably well with the RMT predictions at lower and middle scales. This points
out that the spacing distribution can shed considerable light on the nature of
the high frequency fluctuation correlation, for a smaller corpus of data. 
\par
In conclusion, wavelet analysis when combined with RMT approach can reveal
considerable information about the correlations at different scales and is quite
useful for modeling the behavior of high and low frequency components of
physical processes.
\label{sec:conc}
\bibliographystyle{spphys}
\bibliography{manuscript3.bbl}
\end{document}